\documentclass[aps,prb,reprint,twocolumn,superscriptaddress]{revtex4-2}

\usepackage{graphicx}
\usepackage{hyperref}
\usepackage{amsmath}
\usepackage{siunitx}
\usepackage{color}
\usepackage{float}

\begin{document}
%\preprint{}

\title{Thermal transport of the frustrated spin-chain mineral linarite:\\ Magnetic heat transport and strong spin-phonon scattering}


\author{Matthias Gillig}
\email[]{m.gillig@ifw-dresden.de}
\affiliation{Leibniz Institute for Solid State and Materials Research, 01069 Dresden, Germany}
\affiliation{Institute for Solid State Physics, TU Dresden, 01062 Dresden, Germany}

\author{Xiaochen Hong}
\affiliation{Leibniz Institute for Solid State and Materials Research, 01069 Dresden, Germany}
\affiliation{Fakultät für Mathematik und Naturwissenschaften, Bergische Universität Wuppertal, 42097 Wupptertal, Germany}

\author{Piyush Sakrikar}
\affiliation{Indian Institute of Science Education and Research (IISER) Mohali, Knowledge City, Sector 81, Mohali 140306, India}

\author{Ga\"el Bastien}
\author{A.U.B. Wolter}
\affiliation{Leibniz Institute for Solid State and Materials Research, 01069 Dresden, Germany}

\author{Leonie Heinze}
\affiliation{Institut für Physik der Kondensierten Materie, TU Braunschweig, 38106 Braunschweig, Germany}

\author{Satoshi Nishimoto}
\affiliation{Leibniz Institute for Solid State and Materials Research, 01069 Dresden, Germany}
\affiliation{Department of Physics, Technical University Dresden, 01069 Dresden, Germany}

\author{Bernd B\"uchner}
\affiliation{Leibniz Institute for Solid State and Materials Research, 01069 Dresden, Germany}
\affiliation{Institute for Solid State Physics, TU Dresden, 01062 Dresden, Germany}
\affiliation{Center for Transport and Devices, TU Dresden, 01062 Dresden, Germany}

\author{Christian Hess}
\affiliation{Leibniz Institute for Solid State and Materials Research, 01069 Dresden, Germany}
\affiliation{Fakultät für Mathematik und Naturwissenschaften, Bergische Universität Wuppertal, 42097 Wupptertal, Germany}
\affiliation{Center for Transport and Devices, TU Dresden, 01062 Dresden, Germany}
%\thanks{}


\date{\today}

\begin{abstract}

The mineral linarite (PbCuSO$_4$(OH)$_2$) forms a monoclinic structure where a sequence of Cu(OH)$_2$ units forms a spin-$\frac{1}{2}$ chain. 
Competing ferromagnetic nearest-neighbor ($J_1$) and antiferromagnetic next-nearest-neighbor interactions ($J_2$) in this quasi-one-dimensional spin structure imply magnetic frustration and lead to magnetic ordering below $T_N = \SI{2.8}{\kelvin}$ in a mutliferroic elliptical spin-spiral ground state.
Upon the application of a magnetic field along the spin-chain direction, distinct magnetically ordered phases can be induced.
We studied the thermal conductivity $\kappa$ in this material across the magnetic phase diagram as well as in the paramagnetic regime in the temperature ranges \SIrange{0,07}{1}{\kelvin} and \SIrange{9}{300}{\kelvin}.
We found that in linarite the heat is carried mainly by phonons but shows a peculiar non-monotonic behavior in field. In particular, $\kappa$ is highly suppressed at the magnetic phase boundaries, indicative of strong scattering of the phonons off critical magnetic fluctuations.
Even at temperatures far above the magnetically ordered phases, the phononic thermal conductivity is reduced due to scattering off magnetic fluctuations.
The mean free path due to spin-phonon scattering ($l_{\text{spin-phonon}}$) was determined as function of temperature. 
A power law behavior was observed mainly above \SI{500}{\milli \kelvin} indicating the thermal activation of spin fluctuations.
In the critical regime close to the saturation field, $l_{\text{spin-phonon}}$ shows a $1/T$ dependence.
Furthermore, a magnon thermal transport channel was verified in the helical magnetic phase. We estimate a magnon mean free path which corresponds to about 1000 lattice spacings.

\end{abstract}

\pacs{}
%\keywords{}

\maketitle

%\section{}
%\subsection{}
%\subsubsection{}

%%%%%%%%%%%%%%%%%%%%%%%%%%%%%%%%%%%%%%%%%%%%%%%%%%%%%%%%%%%%%%%%%%%%%%%%%%%%%%%%%%%%%%%%%%%%%%%%%%%%%%%%%%%%%%%%%%%%%%%%%%%%%%%%

\section{Introduction \label{sec:Introduction}}
%	description and importance of general field; state of the art; unsolved problems, new work done by the authors
	Low-dimensional spin systems are of interest because of the expected occurrence of strong quantum effects.
	Particularly, spin chains attract attention due to the possibility to host novel ground states, \textit{e.g.}, Tomonaga-Luttinger liquids.
	Heat transport experiments in such systems can elucidate the characteristics by probing the emerging spin excitations. 
	For isotropic spin-$\frac{1}{2}$ Heisenberg chains a finite thermal Drude weight has been predicted due to the conservation of the heat current \cite{Zotos.1997, Zotos.1999, Kluemper.2002} which renders the magnetic heat transport ballistic.
	Indeed, in material realizations of the isotropic spin-$\frac{1}{2}$ Heisenberg chain, such as the cuprate compounds SrCuO$_2$ and Sr$_2$CuO$_3$, experimental evidence for ballistic spinon heat transport has been inferred from highly anisotropic heat conductivity tensors \cite{Kawamata.2008, Hlubek.2010, Hlubek.2012, Hess.2019}.
	Recently, a non-zero Drude weight was predicted also in a frustrated spin chain, where the interaction among nearest and next-nearest neighbors are competing \cite{Stolpp.2019}. Experimentally, except a pioneering study in LiCu$_2$O$_2$, where a magnon heat channel below the ordering temperature was observed \cite{Liu.2011}, heat transport data of frustrated spin-$\frac{1}{2}$ Heisenberg chains is scarce.
	
	In this paper we present heat transport experiments on the frustrated spin-chain material linarite (PbCuSO$_4$(OH)$_2$).  
	It is a well studied natural mineral which crystallizes in a monoclinic lattice (space group: $P2_1/m$) \cite{Effenberger.1987,Schofield.2009} where a sequence of Cu(OH)$_2$ units forms a quasi-one-dimensional structure of Cu$^{2+}$ ions along the $b$-direction. Due to the spacial separation of these strips along $a$ and $c$ it can be described as a quasi-1D spin chain with $S=\frac{1}{2}$.
	The magnetic interaction between the spins has been modeled based on different experimental results (neutron scattering \cite{Rule.2017}, magnetic susceptibility \cite{Wolter.2012, Schaepers.2013}) which render competing exchange interactions between nearest neighbors ($J_1 \approx \SI{-78}{\kelvin}$) and next-nearest neighbors ($J_2 \approx \SI{28}{\kelvin}$) \cite{Rule.2017}.
	Additional finite interchain couplings lead to long range magnetic order below $T_N = \SI{2.8}{\kelvin}$ where the spins are arranged in an helical incommensurate state with the propagation vector $\vec{q} = (0; 0.186; 0.5)$ \cite{Willenberg.2012}. This magnetic ordering implies spin-driven ferroelectricity unveiling strong magnetoelastic coupling in linarite \cite{Mack.2017}.
	Given the estimated Curie-Weiss temperature $\Theta_{CW} = \SI{27}{\kelvin}$ \cite{Wolter.2012}, the frustration ratio $f = \frac{\Theta_{CW}}{T_N} \approx 10$ points out the high degree of frustration in the system.
	The magnetic phase diagram strongly depends on the direction of the magnetic field \cite{Cemal.2018,Feng.2018} which underlines the anisotropic character of the magnetic system. Here, we focus on the particular case $\mathbf{H}||\mathbf{b}$, which is the best studied case up to date.
	
\begin{figure}[h]
		\includegraphics[width= 0.9 \linewidth]{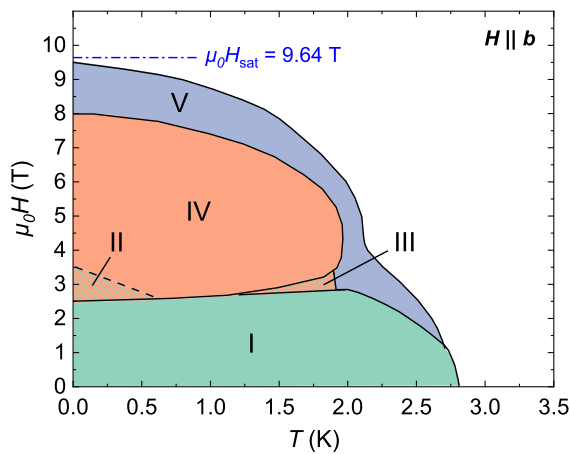}
		\caption{Magnetic phase diagram of PbCuSO$_4$(OH)$_2$ for $\mathbf{H}||\mathbf{b}$ adapted from \cite{Heinze.2019}. The Roman numerals label the magnetic phases: helical incommensurate phase (I), hysteretic commensurate region (II), coexistence phase (III), collinear Néel antiferromagnetic order (IV), and spin-density wave phase (V).\label{fig:PhaseDiagram}}
	\end{figure}

	Upon the application of a magnetic field along $b$, several transitions into distinct magnetic phases are induced \cite{Schaepers.2013, Povarov.2016, Willenberg.2012, Willenberg.2016}. They are depicted in the magnetic phase diagram in figure \ref{fig:PhaseDiagram}. 
	The helical magnetic phase, which at $\mu_0 H=\SI{0}{\tesla}$ sets in at $T_N\approx\SI{2.8}{\kelvin}$, is suppressed by a magnetic field and below \SI{2}{\kelvin} and above \SI{2.5}{\tesla} it gives way to a collinear Néel antiferromagnetic order (IV) with a commensurate propagation vector $\vec{q} = (0; 0; 0.5)$.
	% Spins couple FM along $a$ and $b$, AFM along $c$. Spins lie in $ac$ plane.
	At very low temperature ($T < \SI{700}{\milli \kelvin}$), a hysteretic region II separates phase I from IV with a commensurate vector $\vec{q}=(0;0;0.5)$ whereas above \SI{1}{\kelvin} the phases I and IV are connected via a coexistence phase III where the commensurate and an incommensurate, circular spin arrangement coexist \cite{Willenberg.2016}.
	The mentioned phases are enclosed by a phase which was refined as an incommensurate longitudinal spin-density-wave state \cite{Willenberg.2016} with $\vec{q}=(0; k_y; 0.5)$ which depends on the magnetic field and also on temperature \cite{Heinze.2019}.
	Beyond the saturation field of \SI{9.64}{\tesla}, determined by a nuclear magnetic resonance (NMR) study \cite{Heinze.2019}, the system is field polarized.

%%%%%%%%%%%%%%%%%%%%%%%%%%%%%%%%%%%%%%%%%%%%%%%%%%%%%%%%%%%%%%%%%%%%%%%%%%%%%%%%%%%%%%%%%%%%%%%%%%%%%%%%%%%%%%%%%%%%%%%%%%%%%%%%

\section{Experimental approach \label{sec:ExperimentalApproach}}

	The thermal conductivity $\kappa$ has been measured between \SI{9}{\kelvin} and \SI{300}{\kelvin} in a home-built setup with a standard steady-state method where heat was applied on one side of the sample using a chip resistor whereas the other side of the crystal was mounted on a cold bath. The applied thermal gradient along the sample was measured with a Au-Chromel thermocouple.
	For measurements between \SI{0,07}{\kelvin} and \SI{1}{\kelvin} the thermal gradient was measured with in-situ calibrated RuO$_2$ sensors using a dilution refrigerator. Magnetic fields of up to \SI{16}{\tesla} were applied along the $b$-direction in all measurements. The thermal conductivity measurements as a function of temperature with fixed magnetic field were performed in a steady-state mode, where the given temperature is defined as the average of the temperature at the hot and cold thermometer.
	For field-dependent thermal conductivity measurements the temperature of the heat bath and the heating power were fixed and the magnetic field was ramped slowly with a maximum rate of \SI{0.06}{\tesla \per \minute} to avoid heating effects. 
	One should note that in this measurement mode, due to strong changes in $\kappa$, the temperature of the sample slightly varies while sweeping the field, while for our representation of the results, the average temperature of each field ramp is given.
	We report experimental results from measurements on two different specimens of naturally grown crystals (both from the Grand Reef Mine, Graham County, Arizona, USA). Sample 1 was used to investigate the heat transport along $b$. The shape of sample 2 allowed to inject a heat current perpendicular to $b$. The heat current direction is $-20^{\circ} \pm 5^{\circ}$ off the $a$-axis and was chosen to maximize the length of the crystal along the heat current.

%%%%%%%%%%%%%%%%%%%%%%%%%%%%%%%%%%%%%%%%%%%%%%%%%%%%%%%%%%%%%%%%%%%%%%%%%%%%%%%%%%%%%%%%%%%%%%%%%%%%%%%%%%%%%%%%%%%%%%%%%%%%%%%%

\section{Results \label{sec:Results}}
	\subsection{Thermal conductivity $T \geq \SI{9}{\kelvin}$ (paramagnetic state)}	
		
		\begin{figure}
			\includegraphics[width= \linewidth]{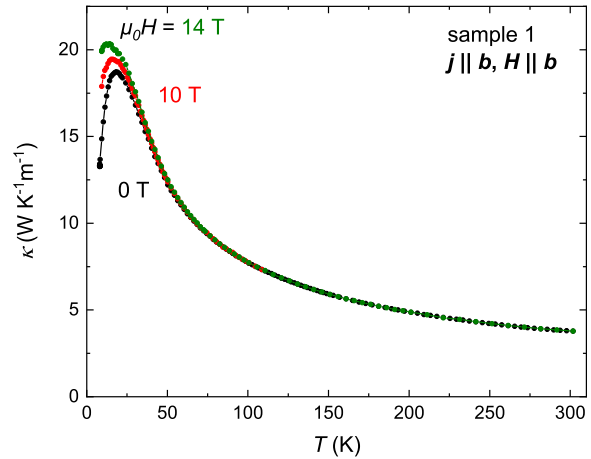}
			\caption{Thermal conductivity for $T > \SI{9}{\kelvin}$ with magnetic field applied along the spin-chain direction ($\mathbf{H}||\mathbf{b}$) \label{fig:kappa_highT}}
		\end{figure}
		
		The results for the thermal conductivity measurements in the temperature range from \SIrange{9}{300}{\kelvin} are shown in figure \ref{fig:kappa_highT}. The zero-field curve resembles that of a phonon heat conductor with a maximum of \SI{18.7}{\watt \per \kelvin\per \meter} at \SI{18}{\kelvin} and a decrease towards higher temperatures.
		Upon applying a magnetic field parallel to the spin-chain direction ($\mathbf{H}||\mathbf{b}$), the peak in $\kappa$ is enhanced below \SI{50}{\kelvin}. At \SI{18}{\kelvin} the increase in field is about \SI{3.4}{\percent} for \SI{10}{\tesla} and \SI{6.8}{\percent} for \SI{14}{\tesla}. 
		Field-enhanced thermal conductivity is not expected for a pure phonon system and in a paramagnetic state it can only be explained within a scenario of magnetic scattering of the phonons at \SI{0}{\tesla} which is gradually reduced in magnetic field, thus leading to the increase in $\kappa$.
		Electron spin resonance experiments \cite{Wolter.2012} show a broadening of resonance lines around \SI{50}{\kelvin} which signals the occurrence of magnetic fluctuations far above the ordering temperature. Since the onset of the fluctuations occurs at the same temperature as the field-enhancement of $\kappa$, they can be hold responsible for the phonon scattering. In this scenario, the decline of magnetic scattering in the $\kappa$-data is consistent with a field-induced reduction of the spin fluctuations.

	\subsection{Thermal conductivity $T\leq\SI{1}{\kelvin}$}	
		The thermal conductivity of sample 1 as function of temperature for $T<\SI{1}{\kelvin}$ is shown in figure \ref{fig:kappaTvsT-sample1-square} with the heat current and the magnetic field applied along the spin-chain axis.
		The zero-field thermal conductivity increases monotonically from \SI{75}{\milli \kelvin} to \SI{1}{\kelvin}. $\kappa$ exhibits a temperature dependence that scales with $\propto T^{2.50}$ ($T<\SI{0,4}{\kelvin}$) and $\propto T^{2.04}$ ($\SI{0,4}{\kelvin}<T<\SI{1}{\kelvin}$) (see fig. \ref{fig:kappa_fit_0T}) which deviates from the standard phonon expression $\kappa \propto c_V \propto T^3$.
		In this temperature and field regime the spin system orders in an elliptical spiral, so it is expectable  that both phonons and magnons contribute to the thermal conductivity directly and at the same time act as a potential scatterer of the respective other quasiparticle.
		Since neither the temperature dependence nor the absolute value of either contribution is known, it is not straight forward to disentangle the two contributions.
		
		To this end the application of a magnetic field can help to resolve this since it should not affect a purely phononic heat conduction channel.
		Figures \ref{fig:kappaTvsT-sample1-square} (a)-(c) depict how the temperature dependence of the thermal conductivity evolves under magnetic field in the regimes of \SIrange{0}{4}{\tesla}, \SIrange{4}{9}{\tesla} and \SIrange{9}{16}{\tesla}, respectively.
		In the lowest range (fig. \ref{fig:kappaTvsT-sample1-square} (a)) and below \SI{0.6}{\kelvin}, $\kappa$ is only weakly altered by a magnetic field. However, with increasing temperature the field effect intensifies. Between \SIrange{0.6}{1}{\kelvin} it is clearly noticeable that $H$ suppresses $\kappa$ with the strongest impact at \SI{2.5}{\tesla} where it is diminished by \SI{29}{\percent} at \SI{1}{\kelvin}. With further increase of $H$, $\kappa$ recovers about \SI{50}{\percent} of the suppressed fraction at \SI{4}{\tesla}. 
		Above \SI{6}{\tesla} (fig. \ref{fig:kappaTvsT-sample1-square} (b)) another decrease sets in, diminishing $\kappa$ first at higher temperature, and for $\mu_0H>\SI{7.5}{\tesla}$ the region of suppressed $\kappa$ is subsequently shifted  towards lower temperature with increasing magnetic field. The transition between these two regions is clearly visible and traces the phase boundary in the $H$-$T$ phase diagram between phase IV and V. Below \SI{200}{\milli \kelvin} the field induced effects are reduced.
		The graph in panel \ref{fig:kappaTvsT-sample1-square} (c) shows the results for $\mu_0H>\SI{9}{\tesla}$ where the trend in the field dependence is inverted and $\kappa$ recovers quickly until it becomes field independent at \SI{12}{\tesla} and higher.
		Over the whole field range, a strong change in the temperature dependence of $\kappa$ is observed, the most pronounced changes occur when the $\kappa(T)$ curve crosses a phase boundary, \textit{e.g.}, at \SI{8.1}{\tesla} exhibiting a steep slope below \SI{0.5}{\kelvin} whereas for $T>\SI{0,7}{\kelvin}$ the increase with temperature is smaller.
		%%%%%%%%%%%%%%%%%%%%%%%
		
		At \SI{16}{\tesla}, far above the saturation field, it is expected that all magnetic excitations are gapped out and the low-temperature thermal conductivity in this regime is independent of the magnetic system and thus purely phononic.
		Comparing this to the smaller $\kappa$ in field below \SI{16}{\tesla}, it is obvious that the thermal transport is dominated by phonons and the respective reduction must be assigned to spin-phonon scattering. Measurements of the thermal expansion have also shown previously that a strong coupling of the lattice to the magnetic system is present in linarite \cite{Schaepers.2013}. Magnetic heat transport can, however, not be ruled out, but it would require a detailed knowledge of the magnetic excitations to disentangle their intricate interplay with the phonon system.
		However, at \SI{0}{\tesla}, the thermal conductivity is enhanced compared to the phononic background. This can only be attributed to an additional magnetic contribution $\kappa_{\text{mag}}$ to the thermal conductivity.
		
	\subsubsection*{Anisotropy of $\kappa$ - evidence for magnon heat transport in the helical phase}
		
\begin{figure*}
			\includegraphics[width=0.8\linewidth]{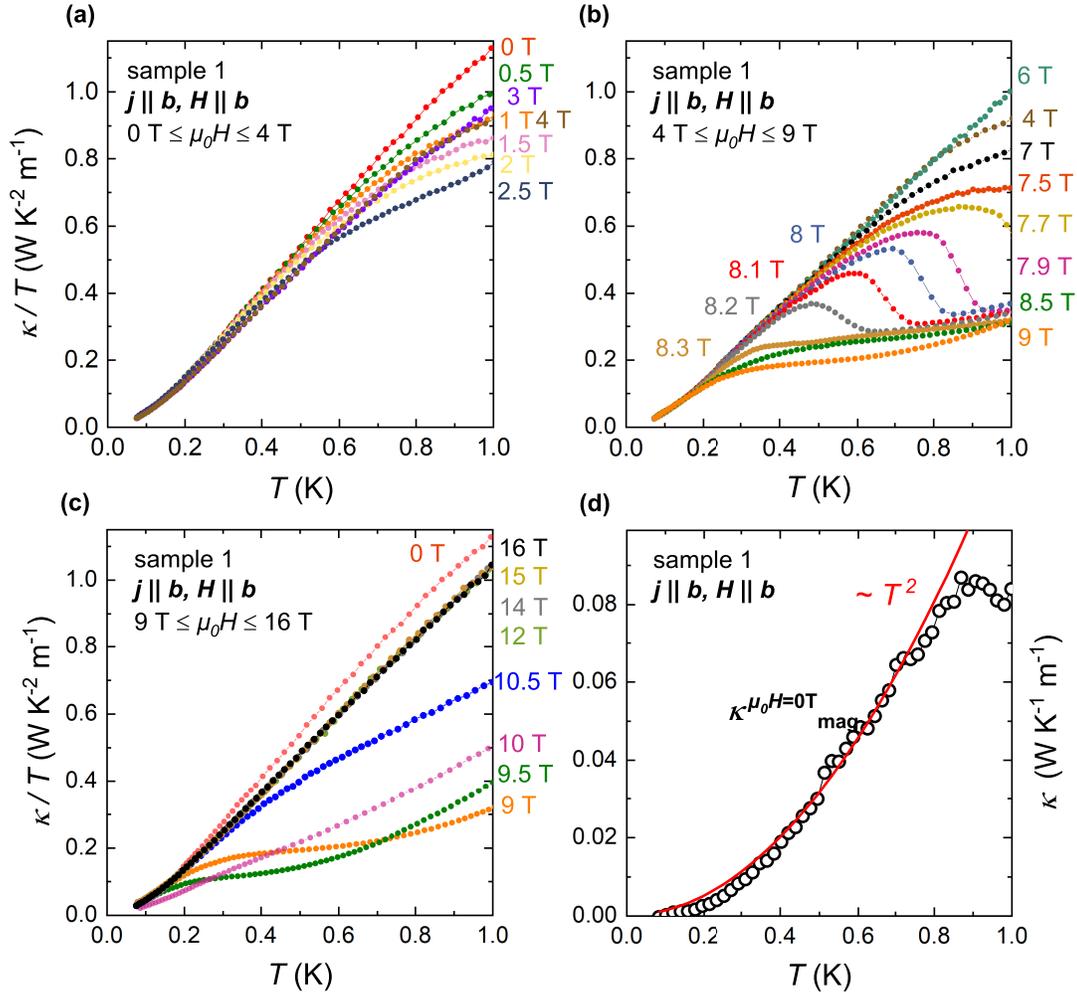}
			\caption{Temperature dependence of the low-temperature thermal conductivity of sample 1 with the heat current and magnetic field oriented along the $b$-direction for the magnetic field ranges \textbf{(a)} 0-4 T, \textbf{(b)} 4-9 T and \textbf{(c)} 9-16 T. In \textbf{(d)} the open circles depict the magnetic contribution to the thermal conductivity at zero field obtained from subtracting the phononic background in the polarized phase at 16 T from the zero-field curve $\kappa_{\text{mag}}^{0\text{T}}=\kappa(0\text{T})-\kappa(16\text{T})$, the solid line is a guide to the eye. \label{fig:kappaTvsT-sample1-square}}
		\end{figure*}

		In order to estimate $\kappa_{\text{mag}}$,  in figure \ref{fig:kappaTvsT-sample1-square} (d) the difference between the zero-field and \SI{16}{\tesla} curve is plotted as function of temperature ($\kappa_{\text{mag}} = \kappa(\SI{0}{\tesla}) - \kappa(\SI{16}{\tesla}))$. $\kappa_{\text{mag}}$ adds \SI{10}{\percent} of the phononic background to the total thermal conductivity and it follows a quadratic temperature dependence up to \SI{0.8}{\kelvin} where a downturn sets in. 
		Due to the strong magneto-elastic coupling across the differently ordered phases, the intense phonon scattering off magnetic excitations must also be considered at \SI{0}{\tesla}, diminishing $\kappa_{\text{ph}}(\SI{0}{\tesla})$ compared to the phononic background in the polarized state. Therefore, at this stage of the analysis the calculated $\kappa_{\text{mag}}$ should be considered a lower bound for the magnetic contribution.
				
		%\onecolumngrid	
		
%		\twocolumngrid
		
\begin{figure*}
			\includegraphics[width=0.8 \linewidth]{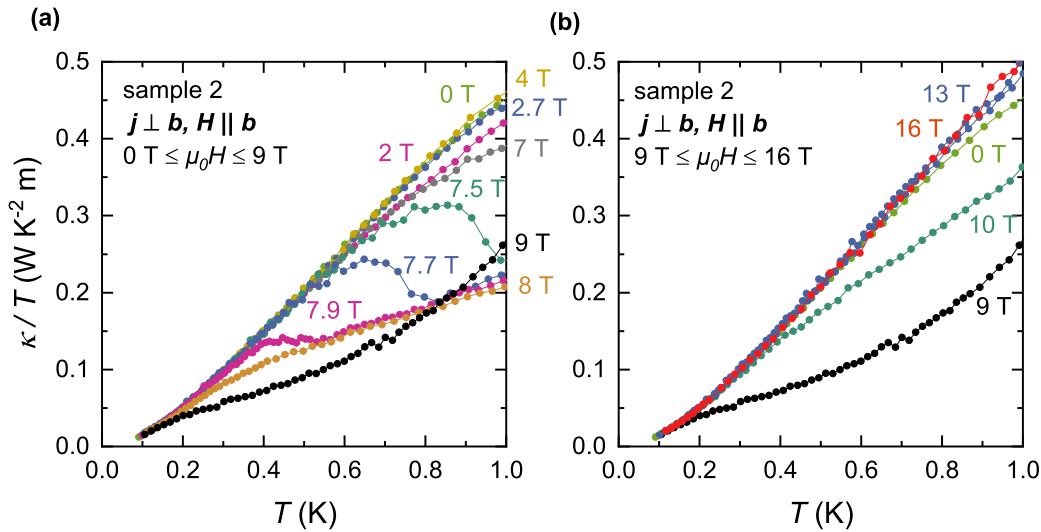}
			\caption{Temperature dependence of the low-temperature thermal conductivity of sample 2  with the heat current perpendicular and the magnetic field parallel to the $b$-direction for the magnetic field ranges \textbf{(a)}  0-9 T and \textbf{(b)} 9-16 T.\label{fig:kappaTvsT-sample2}}
		\end{figure*}

		%%% SAMPLE 2
		In order to clarify to what extent magnetic phonon scattering must be considered to extract the correct value of $\kappa_{\text{mag}}$, $\kappa$ has been measured on an additional sample 2 (see fig. \ref{fig:kappaTvsT-sample2}) along a direction perpendicular to the spin-chain axis. Due to the anisotropy of the magnetic lattice, no significant magnetic heat transport is expected, whereas the phonon contribution and its scattering behavior is assumed to be isotropic.
		Figure \ref{fig:kappaTvsT-sample2} (a) shows the results for $\mu_0 H\leq\SI{9}{\tesla}$. $\kappa/T$ differs from sample 1 by a fixed factor of about two, while the temperature dependence and the systematic change in magnetic field are the same. 
		The different absolute value can be explained by differences in the geometric factor of the two samples that were not perfectly defined due to the natural grown shape of the crystals. 
		On the other hand, the apparent identical field dependence confirms the validity of the assumption of isotropic phonon scattering.
		In this configuration the thermal conductivity traces the features of the phase diagram by exhibiting significant changes close to a magnetic phase transition.
		In panel (b), the thermal conductivity for $\mu_0 H\geq\SI{9}{\tesla}$ is depicted. At \SI{9}{\tesla} it is suppressed most prominently before it recovers at a larger field and becomes field independent at \SI{13}{\tesla} and above. Most importantly, the zero-field curve matches vastly with the phononic background at \SI{16}{\tesla} which means that there is no detectable additional contribution at zero field compared to the phononic background. Furthermore there is also not a significant reduction of $\kappa/T(\SI{0}{\tesla})$ due to magnetic scattering, merely above \SI{0.7}{\kelvin} a slight deviation sets in. Consequently, $\kappa_{\text{mag}}$ determined in sample 1 indeed is the actual magnetic contribution of the spiral spin system to heat transport up to \SI{0.7}{\kelvin}. 
		Above \SI{0,7}{\kelvin}, phonon scattering sets in also at \SI{0}{\tesla}. Since $\kappa(\SI{0}{\tesla})$ is reduced below the phononic background, the simple subtraction does not yield a meaningful $\kappa_{\text{mag}}$ in this temperature range. This gives an explanation for the kink-like feature that appears in $\kappa_{\text{mag}}$ at \SI{0,9}{\kelvin} in figure \ref{fig:kappaTvsT-sample1-square} (d) where an overestimation of the phonon contribution produces such an artifact.
		Also for the $\kappa(T)$ curves for fields $\mu_0 H>\SI{0}{\tesla}$, where $\kappa$ is below the phononic background, it is not straightforward to disentangle possible magnon contributions from phonons since both quasiparticles are expected to scatter off each other which suppresses the respective fraction of $\kappa$.
		
	\subsubsection*{Field dependence of $\kappa$ - evidence for strong spin-phonon scattering}
		
\begin{figure}[h]
			\includegraphics[width= \linewidth]{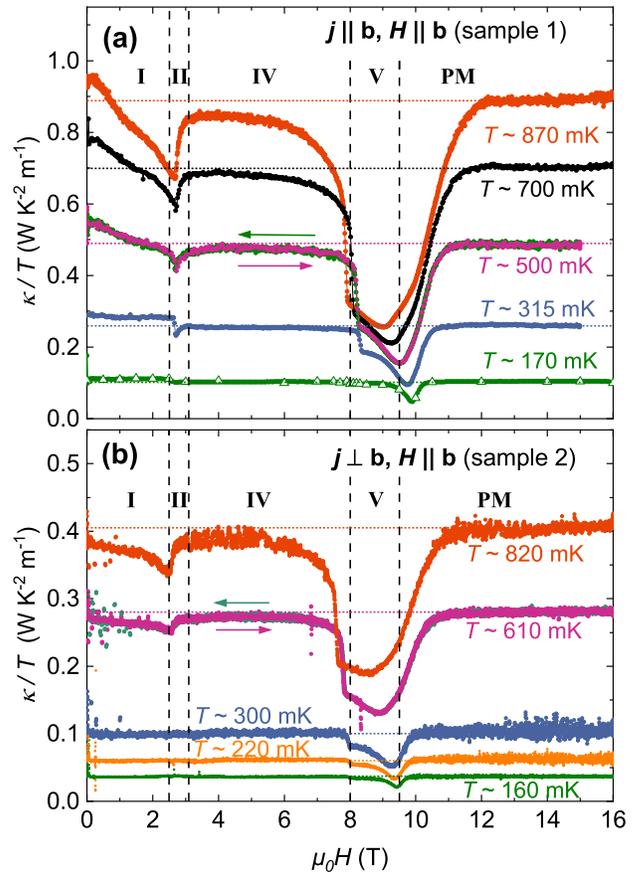}
			\caption{Thermal conductivity as function of magnetic field where in \textbf{(a)} both the heat current and the magnetic field are applied along $b$ in sample 1. \textbf{(b)} shows the thermal conductivity $\kappa/ T$ of sample 2 as a function of magnetic field where the heat current $j$ is applied perpendicular to $b$ and the magnetic field parallel to $b$. The dashed lines correspond to the phase boundaries at $T \approx 250$ mK as referred to in  \cite{Schaepers.2013}. The dotted lines refer to the phononic background which is defined in the fully polarized phase. \label{fig:kappaTvsB}}
		\end{figure}

		Figure \ref{fig:kappaTvsB} shows the thermal conductivity $\kappa/T$ as function of magnetic field at various temperatures for sample 1 $(\mathbf{j} || \mathbf{b})$ in panel (a) and sample 2 $(\mathbf{j} \perp \mathbf{b})$ in panel (b). 
		It provides additional information about abrupt changes in $\kappa$ under magnetic field.
		As mentioned in section II, $\kappa$'s field dependence leads to a relative variation of the sample temperature which is however smaller than the relative changes in $\kappa$ itself. 
		This effect is most significant in phase V where the mean temperature of the sample as compared to phase I is raised by \SI{10}{\percent}. 
		Despite this variation of the sample temperature, the data for $\kappa(H)$ are well consistent with those for $\kappa(T)$ where the temperature variation is absent. This can be verified in fig. \ref{fig:kappaTvsB} (a), \textit{e.g.}, for the data at $T \approx \SI{170}{\milli\kelvin}$  where the triangles represent data from constant field measurements.
		
		For the heat transport along the spin chain direction ($\mathbf{j}||\mathbf{b}$) (fig. \ref{fig:kappaTvsB} (a)), the field dependence is similar for all temperatures. Compared to the high-field limit, which defines the maximum phononic background, $\kappa$ is mostly reduced at lower fields. Only in phase I, $\kappa$ exceeds the high-field limit due to the additional magnetic contribution. At lowest temperature, $\kappa$ seems largely field independent. However at the phase transitions sharp features occur. From phase I to phase II there is a step-like change which corresponds to switching off the magnetic transport channel. At higher temperatures, this step is accompanied by a dip implying enhanced phonon scattering at the critical field value. The proximity to ferroelectric order in phase I \cite{Mack.2017} might also affect the phonon transport there.
		The phase line between II and IV is not represented by a sharp feature in the thermal conductivity data. 
		At \SI{8,3}{\tesla} another step down indicates the onset of the spin density wave phase (V) at which the \SI{315}{\milli \kelvin} curve is reduced by about \SI{20}{\percent}. With increasing temperature this abrupt change intensifies in the relative height. Also at higher $T$, a slow decrease of $\kappa$ sets already in at lower fields within phase IV. 
		Beyond the step, the strong dip becomes even deeper until a minimum is reached which coincides with the upper phase boundary of region V. For higher fields, the thermal conductivity recovers steadily to the high-field limit over a range of \SI{0.5}{\tesla} at \SI{170}{\milli \kelvin} to up to \SI{3}{\tesla} at \SI{870}{\milli\kelvin}. The \SI{500}{\milli \kelvin} curve was measured for ramping the magnetic field up and down without signatures for hysteretic behavior in any of the features within experimental resolution. All the distinct features and the explicit field dependence underline the impact of the magnetic excitations on the phonon system. Especially at the phase boundaries where an intensification of magnetic fluctuations is expected the phonons are scattered most strongly. Compared to previous experimental findings \cite{Willenberg.2012, Schaepers.2013, Povarov.2016,Feng.2018,Heinze.2019}, the features in $\kappa(H)$ identified as phase boundaries are slightly higher on the axis of magnetic field ($<\SI{5}{\percent}$), which might be due to sample dependency or marginal misalignment of the $b$-axis with magnetic field.
					
		The graph in panel \ref{fig:kappaTvsB} (b) depicts the field dependence of $\kappa/T$ perpendicular to the spin-chain direction, measured on sample 2. The overall behavior agrees with the result of sample 1. $\kappa (H)$ being largely isotropic validates the phononic predominance of the thermal conductivity. The clear difference between the two orientations is the absence of the contribution in phase I which exceeds the phononic background and is attributed to magnetic heat transport in the spin chain. Also the effect of a magnetic field on this magnetic channel can be determined qualitatively from figure \ref{fig:kappaTvsB} (a).  Above \SI{500}{\milli\kelvin}, $\kappa_{\text{mag}}$ decreases approximately linearly with increasing field, whereas for the two curves below \SI{500}{\milli \kelvin} $\kappa_{\text{mag}}$ seems to form a field-independent plateau across phase I.

\begin{figure}[h]
			\centering
			\includegraphics[width = 1 \linewidth]{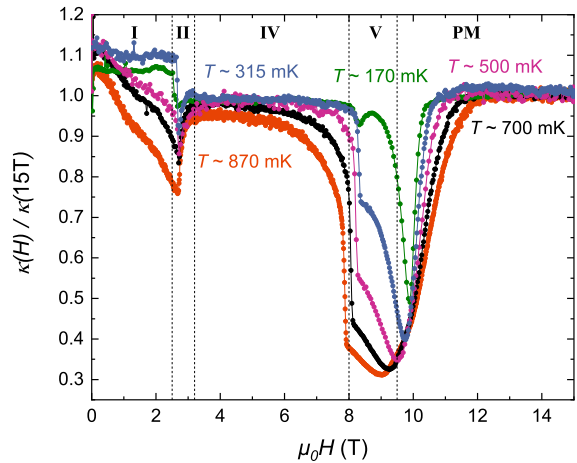}
			\caption{Relative field dependence of thermal conductivity in sample 1 ($\textbf{j, H} || \textbf{b}$) where $\kappa$ is normalized to the value at \SI{15}{\tesla} in the polarized phase. The dashed lines refer to the phase boundaries at \SI{250}{\milli \kelvin} as referred to in \cite{Schaepers.2013}. \label{fig:MC-sample1}}
		\end{figure}

		Figure \ref{fig:MC-sample1} depicts the relative field dependence in sample 1 where $\kappa$ is normalized to the value at 15 T. This highlights the temperature dependence of the features in $\kappa$ that occur while sweeping the field. The most pronounced changes occur in phase I and V. Below 3 T, the field independence softens with increasing temperature, leading to a decline of $\kappa$ towards the phase boundary.
		An also peculiar behavior occurs in phase V and in the vicinity of its phase boundaries. 
		The sharp step that marks the phase transition at \SI{7,9}{\tesla} and \SI{870}{\milli \kelvin} becomes weaker with lowering the temperature until there is only a small dip left at \SI{170}{\milli \kelvin}. The minimum that coincides with the upper phase boundary also becomes less intense upon cooling. However there is still a deep trench remaining at lowest temperature. Across the whole temperature range a substantial field dependence of $\kappa$ is present between the two phase boundaries.

%%%%%%%%%%%%%%%%%%%%%%%%%%%%%%%%%%%%%%%%%%%%%%%%%%%%%%%%%%%%%%%%%%%%%%%%%%%%%%%%%%%%%%%%%%%%%%%%%%%%%%%%%%%%%%%%%%%%%%%%%%%%%%%%

\section{Discussion \label{sec:Discussion}}
	\subsection{Phonon transport}
	The findings of the above presented experiments show deviations of a simple Debye-like thermal conductor in several aspects which are discussed in the following.
	Across the magnetically ordered phases, $\kappa$ exhibits a strong magnetic field dependence where it is largely reduced below the undisturbed phonon contribution at the high-field limit. Therefore, significant magneto-elastic coupling between phonons and magnetic degrees of freedom must be present which constrains the phonon mean free path. 
	The effect is strongest between 7 and \SI{10,5}{\tesla} which corresponds to phase V and its vicinity. For the phases at lower field, magnetic scattering only becomes relevant above \SI{500}{\milli\kelvin} (see fig. \ref{fig:kappaTvsT-sample2} (a)).
	
\begin{figure}[h]
		\includegraphics[width=\linewidth]{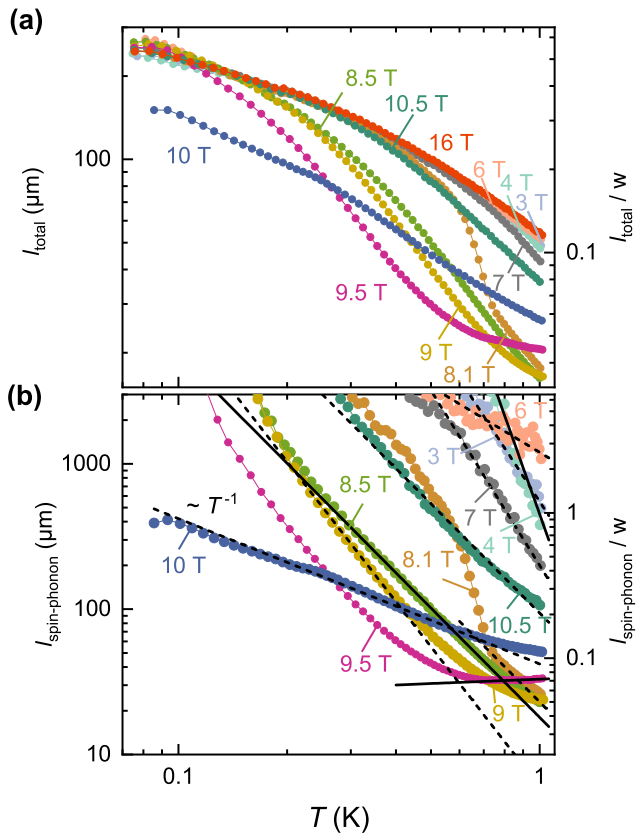}
		\caption{\textbf{(a)} The phonon mean free path $l_{\text{total}}$ as function of temperature in sample 1, \textbf{(b)} spin-phonon mean free path $l_{\text{spin-phonon}}$ which describes the limit of phonon propagation due to spin-phonon scattering. Dashed and solid black lines are guides to the eye and illustrate the power law behavior. The respective exponents are shown in table \ref{tab:powerlaw}. The left scales are normalized to the width of the sample.}
		\label{fig:PhononMFP}
	\end{figure}

	We begin our discussion with the purely phononic part of the thermal conductivity (e.g. at \SI{16}{\tesla}). It shows a deviation from the standard $T^3$ behavior given by the specific heat of the phonon system (see fig. \ref{fig:kappa_fit_16T}). In the low temperature limit of pure phonon systems, the phonon wavelength becomes so large that boundary scattering is the only relevant process to limit the mean free path. It has been shown before that a smooth crystal surface causes phonons to specularly reflect at the boundaries leading to a behavior of $\kappa_{\text{ph}} \propto T^{\alpha}$ where $\alpha = 2-3$ (e.g. in Al$_2$O$_3$ \cite{Pohl.1982} and in LiF \cite{Thacher.1967}), as is also observed in our data, implying  a temperature dependent phonon mean free path.
	In the following we will estimate the phonon mean free path $\l_{ph}$ using the formula
	
	\begin{equation}
		\kappa_{\text{ph}} = \frac{1}{3} \left\langle v\right\rangle  c_V l, % = \frac{1}{3} v_{\text{s}} \frac{\rho}{M} \beta T^3 l ,
		\label{eq:kappa_phonon}
	\end{equation}
	
	with $\left\langle v\right\rangle$ the mean phonon velocity averaged over all branches, $c_V$ the specific heat of the phonon system normalized to the volume, and $l$ the phonon mean free path.
	At low temperature, one can assume that only acoustic phonons contribute with a $k$-independent $v_s$. In this approximation the velocity is given as
	$v_s = \frac{\Theta_D  k_B}{\hbar (6  \pi^2  N )^{1/3}}$ where $\Theta_D$ is the Debye temperature and $N$ the number of elementary cells per unit volume.
	The volumetric specific heat is given as the product of the density of the material $\rho$ and the specific heat of the phonon system. The phononic contribution to the specific heat and the Debye temperature  were determined from a fit of the measured specific heat by a harmonic model in a previous publication \cite{Schaepers.2013}. The Debye temperature is estimated at \SI{133}{\kelvin}. At low temperature the volumetric specific heat is thus estimated with $c_{V} = \frac{\rho}{M} C_{\text{ph}} = \frac{\rho}{M} \beta T^3$ where $M$ stands for the material's molar mass and $\beta = \SI{1.508}{\milli \joule \per \mol \kelvin^4}$.
	Using the mentioned parameters the phonon mean free path is given as
	\begin{equation}
		l_{\text{total}} = \frac{3 \kappa}{c_V v_s}.
		\label{eq:totalMFP}
	\end{equation}
	The temperature dependence of $l_{\text{total}}$ was determined for the data measured on sample 1 and is plotted in figure \ref{fig:PhononMFP} (a) for different magnetic field values.
	$l_{\text{total}}$ at \SI{16}{\tesla} represents the magnetically undisturbed phonon scattering behavior. At \SI{1}{\kelvin} $l_{\text{total}}=\SI{53}{\micro \meter}$, upon cooling it increases to \SI{246}{\micro \meter} at \SI{75}{\milli \kelvin} which corresponds to \SI{53}{\percent} of the sample width. This renders the phonon system close to the ballistic regime where the mean free path is only limited by the sample's edges.
	The double logarithmic representation shows that it does not follow a clear power law over the entire temperature range. Such a behavior has been reported previously  like in the insulating YBa$_2$Cu$_4$O$_6$ where the curvature of $\kappa/T$ changes as function of temperature \cite{Taillefer.1997}.
	
	In order to describe the magnetic scattering separately, we consider an additional magnetic scattering mechanism in the total mean free path $l_{\text{total}}^{-1} = l_{\text{phonon}}^{-1} + l_{\text{spin-phonon}}^{-1}$, according to Matthiessen's rule.
	This spin-phonon mean free path can be separated from $l_{\text{total}}$ considering that the mean free path of the magnetically undisturbed phonons $l_{\text{phonon}}$ is field independent and is equal to the phonon mean free path in the polarized phase ($l_{\text{phonon}} = l_{\text{total}}(\SI{16}{\tesla})$).
	The spin-phonon mean free path is then given as
	\begin{equation}
		l_{\text{spin-phonon}} = \frac{1}{\frac{c_V v_s}{3 \kappa} - l^{-1}_{\text{phonon}}(\SI{16}{\tesla})}.
		\label{eq:spin-phononMFP}
	\end{equation}
	
	$l_{\text{spin-phonon}}$ as function of temperature is depicted in figure \ref{fig:PhononMFP} (b). It can be determined for the fields where $l_{\text{total}}$ differs clearly from $l_{\text{phonon}}$. The plot ranges up to \SI{3000}{\micro \meter}, beyond this value the data scatters strongly due to small differences between $l_{\text{phonon}}$ and $l_{\text{total}}$ of the respective field and therefore does not allow a reasonable interpretation. Spin-phonon scattering becomes mostly relevant in phase V and its vicinity. For \SI{7}{\tesla} and lower, $l_{\text{spin-phonon}}$ is much larger than $l_{\text{phonon}}$, thus the trivial phonon scattering predominates with respect to the magnetic scattering. On the other hand, above \SI{500}{\milli \kelvin} and for all fields, spin-phonon scattering is present and increasing, indicating that the magnetic scattering is thermally activated.
	In certain regions of the temperature range, all the curves obey a power law behavior which is indicated by the solid and dashed black lines in figure \ref{fig:PhononMFP} (b). The respective exponents of the power law are listed in table \ref{tab:powerlaw} in the Appendix.
	Interestingly, at \SI{10}{\tesla}, close to the critical point where the magnetic field begins to polarize the spins, $l_{\text{spin-phonon}}$ follows a $T^{-1}$ dependence over a wide temperature range.
	It is tempting to interpret this as a fingerprint of the critical scattering close to saturation. However, at present, there is no meaningful interpretation of the power law coefficient. However, we believe that the temperature dependence can be useful to determine the spin-phonon scattering matrix element in future works.
		
	The importance of magnetic scattering in phase V with respect to all the other phases is remarkable.
	Electron spin resonance measurements show a continuous decrease of the resonance frequency with field within phase V and its collapse at the saturation field \cite{Gotovko.2019}. Such observations hint at a fan state in phase V.
	In former NMR experiments on this phase, a broad spectrum was observed superimposed with an additional broad peak which follows the NMR signal of the paramagnetic field and temperature range \cite{Willenberg.2016}. This implies that two different local environments are present in phase V, either in a microscopic coexistence or in a phase separation. Previous neutron diffraction measurements reported a reduction of the integrated intensity upon entering phase V \cite{Willenberg.2016,Heinze.2019} which agrees with both afore mentioned scenarios. 
	Furthermore in inelastic neutron scattering (INS) experiments, the spin wave dispersion in phase V appears to be less intense \cite{INS-data} which can be ascribed to either strong spin fluctuations resulting in a lower ordered magnetic moment or to only a partial detection of the magnetic state due to phase separation. 
	The scenario of phase separation would imply that in phase V there are domain walls present in the crystal. The appearance of domain walls causes additional scattering of the phonons and provides an explanation for the strong suppression of the phononic thermal conductivity compared to the other phases. The temperature evolution of the suppression, as shown in figure \ref{fig:MC-sample1}, follows a trend where the scattering becomes less intense towards lowest temperatures. Considering that the phonon wavelength increases with lowering the temperature, one can expect that at lowest temperature it becomes significantly larger than the magnetic domain size reducing the efficiency of phonons being scattered by domain walls. Thereby the thermal conductivity data agrees with the scenario of magnetic phase separation.
	
	Yet the observed transport properties in phase V do not rule out strongly fluctuating spins and in this scope the heavily reduced $\kappa$ can be explained by intense phonon scattering.
	We point out that magnetic fluctuations obviously remain important also at higher temperatures, since they affect $\kappa$ at least up to \SI{50}{\kelvin}.
		
	\subsection{Magnon transport}
\begin{figure}
		\includegraphics[width=0.8\linewidth]{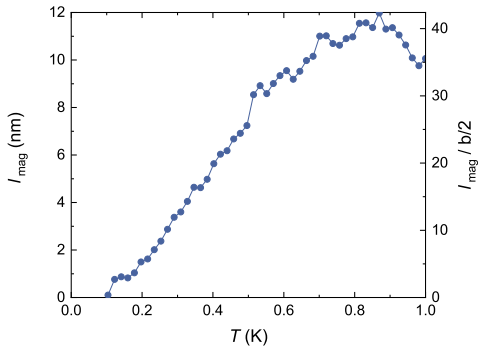}
		\caption{The magnon mean free path of the magnon excitations at $\mu_0 H=\SI{0}{\tesla}$ estimated for a one dimensional system. The right scale is normalized to the distance of magnetic ions.}
		\label{fig:magnonMFP}
	\end{figure}

	A second aspect that distinguishes linarite from a trivial phonon conductor is the magnetic heat transport channel that is present in phase I given by the anisotropic enhancement of $\kappa$ compared to the polarized phase. INS measurements have shown that the dispersion at \SI{0}{\tesla} differs for the two axes $b$ and $c$ \cite{Rule.2017}. Along the chain direction, two minima at $\pm 0.2 \pi/a$ and a crossing of the branches at $0 \pi/a$ were observed. Linear spin wave theory calculations yield a dispersion with maxima of \SI{19}{\milli \electronvolt}.
	A linear fit of the low-energy dispersion yields a magnon velocity of $v_{\text{mag}}^b = \SI{872,5}{\meter \per \second}$ averaged over both branches for positive and negative momenta. The obtained value agrees well with the theoretical modeling of the INS results, yielding a value of $v_{\text{theo}}^b = \SI{883,1}{\meter \per \second}$.
	The measurement along $c$ reveals an energy scale by a factor of 20 smaller with maxima at \SI{1}{\milli \electronvolt}, where the linear fit yields a lower velocity of $v_{\text{mag}}^c = \SI{435}{\meter \per \second}$. Experimental data for dispersions along the $a$-axis have not been reported in the literature so far. However, interactions can safely be assumed negligible given the large interchain distance along the $a$-axis of \SI{9.682}{\angstrom}.
	Due to these anisotropic energy scales the assumption may be valid that a quasi-one-dimensional spin wave excitation carries the heat along the chain whose thermal conductivity is given in the kinetic model by
	\begin{equation}
		\kappa_{\text{mag}} = \frac{n_s l_{\text{mag}} k_B^2 T}{\pi \hbar} \int_{0}^{J \pi / 2 k_B T} \frac{x^2 \exp(x)}{(\exp(x)-1)^2} \text{d}x
	\end{equation}
	
	with $x = \varepsilon_k /k_B T$, the excitation's dispersion function $\varepsilon_k$, a $k$-independent magnon mean free path $l_{\text{mag}}$ and the number of chains per unit area $n_s$.
	For $k_B T \ll |J| = \SI{78}{\kelvin} \cdot k_B$ the integral does not depend on temperature and converges to $\pi^2/3$. With this the magnon mean free path is given as
	\begin{equation}
		l_{\text{mag}}	 =  \frac{3 \hbar}{2 \pi n_s k_B^2} \frac{\kappa_{\text{mag}}}{T}.
	\end{equation}
	The result is presented in figure \ref{fig:magnonMFP} showing a linear temperature dependence up to \SI{0.7}{\kelvin} where it reaches \SI{0,011}{\micro\meter} which corresponds to about 40 chain units. 
	A mean free path that vanishes for $T \rightarrow \SI{0}{\kelvin}$ appears unphysical and challenges the assumptions made.
	
	An alternative method to estimate the magnon mean free path is to use the plain three-dimensional kinetic model with 
	\begin{equation}
		\kappa = \frac{1}{3} c_{\text{V,mag}} v_{\text{mag}} l_{\text{mag}} = \frac{1}{3} C_{\text{mag}} \frac{\rho}{M} v_{\text{mag}} l_{\text{mag}}
		\label{eq:kappa_mag}
	\end{equation}
	which requires the afore mentioned magnon velocity $v_{\text{mag}}^b$ and the magnon contribution to the volumetric specific heat $c_{\text{V,mag} } = \frac{\rho}{M} C_{\text{mag}}$. $C_{\text{mag}}$ has been reported by Schäpers et al. \cite{Schaepers.2013} down to \SI{0.5}{\kelvin}. Below \SI{2}{\kelvin}, it can be described with $C_{\text{mag}} = \gamma T^3$ where $\gamma = \SI{151}{\milli \joule \per \mol \kelvin^4}$.

\begin{figure}
		\includegraphics[width=0.8\linewidth]{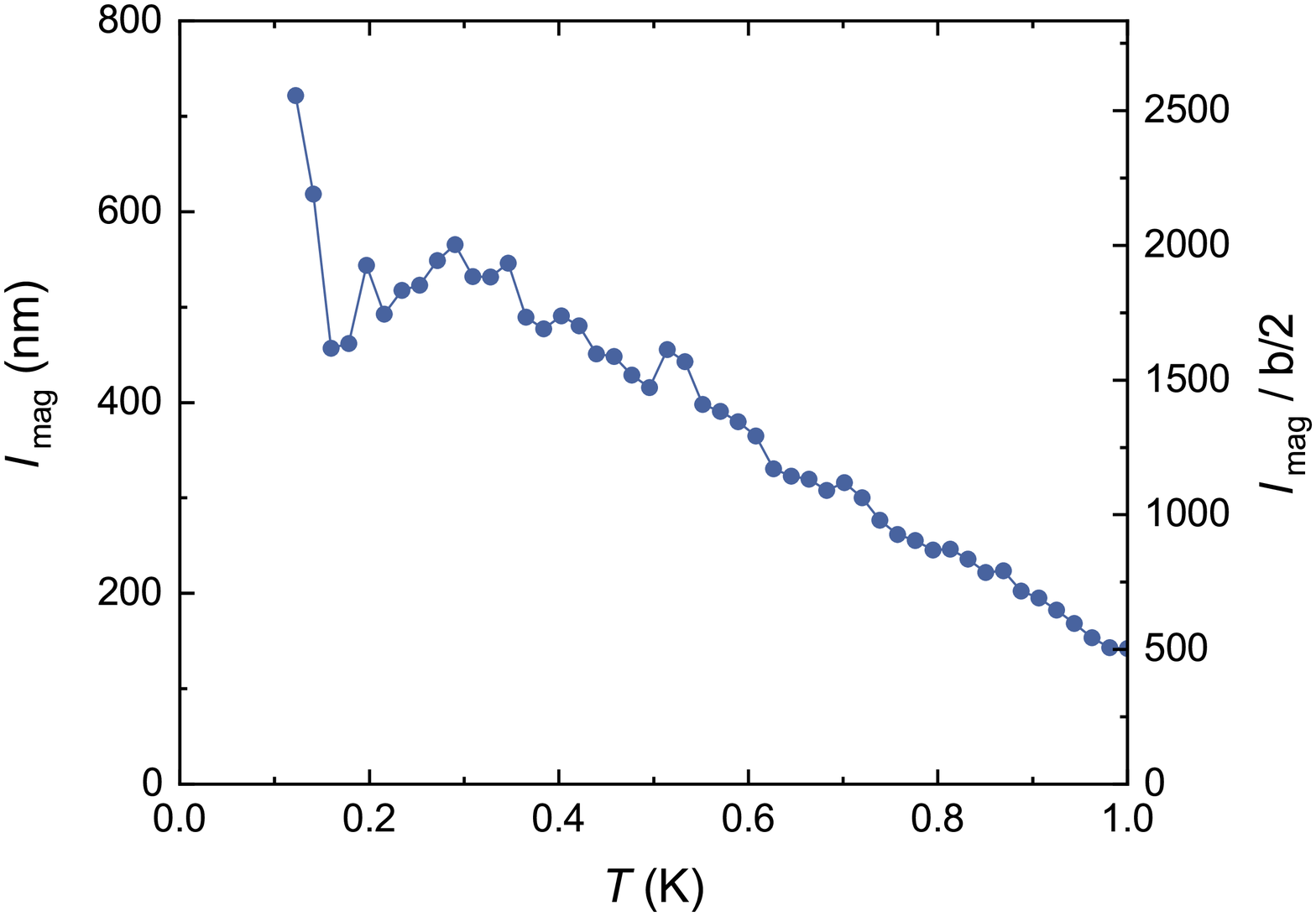}
		\caption{The magnon mean free path of the magnon excitations at $\mu_0 H=\SI{0}{\tesla}$ estimated using the kinetic model with the experimental magnetic specific heat of linarite. The right scale is normalized to the distance of magnetic ions.}
		\label{fig:magnonMFP-exp}
	\end{figure}

	The hereby calculated mean free path is shown in figure \ref{fig:magnonMFP-exp}. It increases towards cooling with a mostly linear slope and reaches \SI{0.5}{\micro \meter} below \SI{0.3}{\kelvin} which corresponds to about 2000 chain units. 
	This second method yields a more plausible temperature dependence of the mean free path and gives a more realistic description of the spin system's heat transport.
	However, the estimation contains uncertainties which are mainly the dimensionality of the spin system contained in the prefactor of $\frac{1}{3}$ in equation \ref{eq:kappa_mag} and the approximation of the magnon velocity. These factors may alter the absolute value of $l_{\text{mag}}$ but do not affect the temperature dependence.
	Compared to the spin-chain compound SrCuO$_2$ ($l_{\text{mag}} \approx \SI{1.5}{\micro\meter}$) \cite{Hlubek.2010}, the mean free path is of the same order of magnitude. On the other hand it is considerably larger than in CaCu$_2$O$_3$ \cite{Hess.2007} ($l_{\text{mag}} = \SI{22}{\angstrom}$). In both materials it has been shown that the mean free path scales with the density of defects within the chains. Transferring these findings to our measurements we can determine that in the here studied crystals the average length of chain segments is about 1000 unit cells.
	
	For the regions beyond phase I, magnetic contributions to the thermal conductivity are either absent or too small to significantly exceed the phonon background. According to expression \ref{eq:kappa_mag}, each magnon conductivity depends on three parameters.
	It has been reported that $C_\text{mag}$ varies with the magnetic field but is not heavily suppressed \cite{Schaepers.2013}. This implies that in these phases either the magnon velocity is diminished due to flattened dispersions or the mean free path is limited due to more intense magnon scattering compared to the helical phase or both simultaneously.
	
	Finally, in contrast to the nonfrustrated spin chains (SrCuO$_2$, Sr$_2$CuO$_3$), there is no indication of ballistic heat transport of spinons in the frustrated spin-chain material linarite. The frustration ratio $\alpha = J_2/J_1$ in linarite is negative, whereas the occurrence of a finite Drude weight has been predicted for $\alpha>0$ \cite{Stolpp.2019}. This raises the question if the nature of the next-nearest neighbor interaction can influence the thermal transport behavior and we hope that the present experimental study can motivate further theoretical investigations on thermal transport in a frustrated ferromagnetic $J_1$-$J_2$ chain.
	
%%%%%%%%%%%%%%%%%%%%%%%%%%%%%%%%%%%%%%%%%%%%%%%%%%%%%%%%%%%%%%%%%%%%%%%%%%%%%%%%%%%%%%%%%%%%%%%%%%%%%%%%%%%%%%%%%%%%%%%%%%%%%%%%%%%%%%		
\section{Conclusion}
	We measured the thermal conductivity of linarite across the magnetic phase diagram below \SI{1}{\kelvin} as well as far above $T_N$ in the paramagnetic regime.
	The heat transport is dominated by phonons. However strong magneto-elastic coupling limits the phonon mean free path and leads to a drastic field dependence which is even observable far above the ordering temperature.
	Due to the significant magnetic scattering of the phonons relevant features of the magnetic phase diagram are indirectly detectable by the heat transport of the lattice.
	An estimation of the phonon mean free path in the low temperature limit renders the phonon heat transport close to the ballistic regime where phonon propagation is only limited by the sample boundaries.
	The phonon mean free path due to spin-phonon scattering has been determined as function of temperature and shows different power law behaviors depending on the magnetic field.
	Furthermore we have found that in the helical magnetic phase there is a direct contribution of magnetic excitations to the thermal conductivity parallel to the spin-chain direction. The mean free path of these excitations gives a measure for the length of the materials spin-chain segments.

%%%%%%%%%%%%%%%%%%%%%%%%%%%%%%%%%%%%%%%%%%%%%%%%%%%%%%%%%%%%%%%%%%%%%%%%%%%%%%%%%%%%%%%%%%%%%%%%%%%%%%%%%%%%%%%%%%%%%%%%%%%%%%%%%%%%%%		
\section*{Acknowledgements}
	We thank Stefan Süllow and Kirrily Rule for fruitful discussions.
	This work has been supported by the Deutsche Forschungsgemeinschaft under Contracts No. WO 1532/3-2 and No. SU 229/9-2, through SFB 1143 (project-id 247310070), as well as the Würzburg-Dresden Cluster of Excellence on Complexity and Topology in Quantum Matter ct.qmat (EXC 2147, project-id 390858490).
	This work has been further supported by the European Research Council (ERC) under the European Union’s Horizon 2020 research and innovation program (Grant Agreement No. 647276-MARS-ERC-2014-CoG) and under the Marie Sklodowska-Curie actions (Grant Agreement No 796048).

\newpage
\appendix
\section*{Supplemental material}

\section{Temperature dependence of $\kappa$}

	\begin{figure}[H]
		\centering
		\includegraphics[width = 0.9 \linewidth]{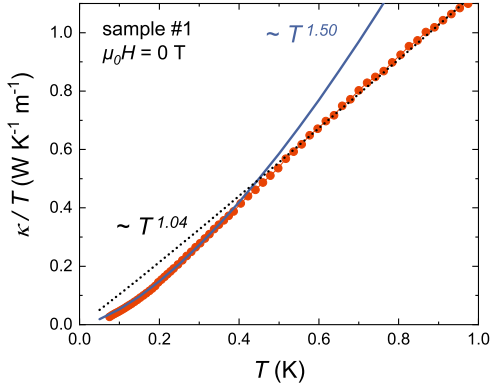}
		\caption{Zero-field thermal conductivity of sample 1 fitted with a power law ($\kappa = C_0 \cdot T^{\alpha}$) where $\alpha = 2.50$ for $T< \SI{0,4}{\kelvin}$ (solid blue line) and $\alpha = 2.04$ for $\SI{0,4}{\kelvin} < T < \SI{1}{\kelvin}$ (dotted black line) \label{fig:kappa_fit_0T}}
	\end{figure}

	\begin{figure}[H]
		\centering
		\includegraphics[width = 0.9 \linewidth]{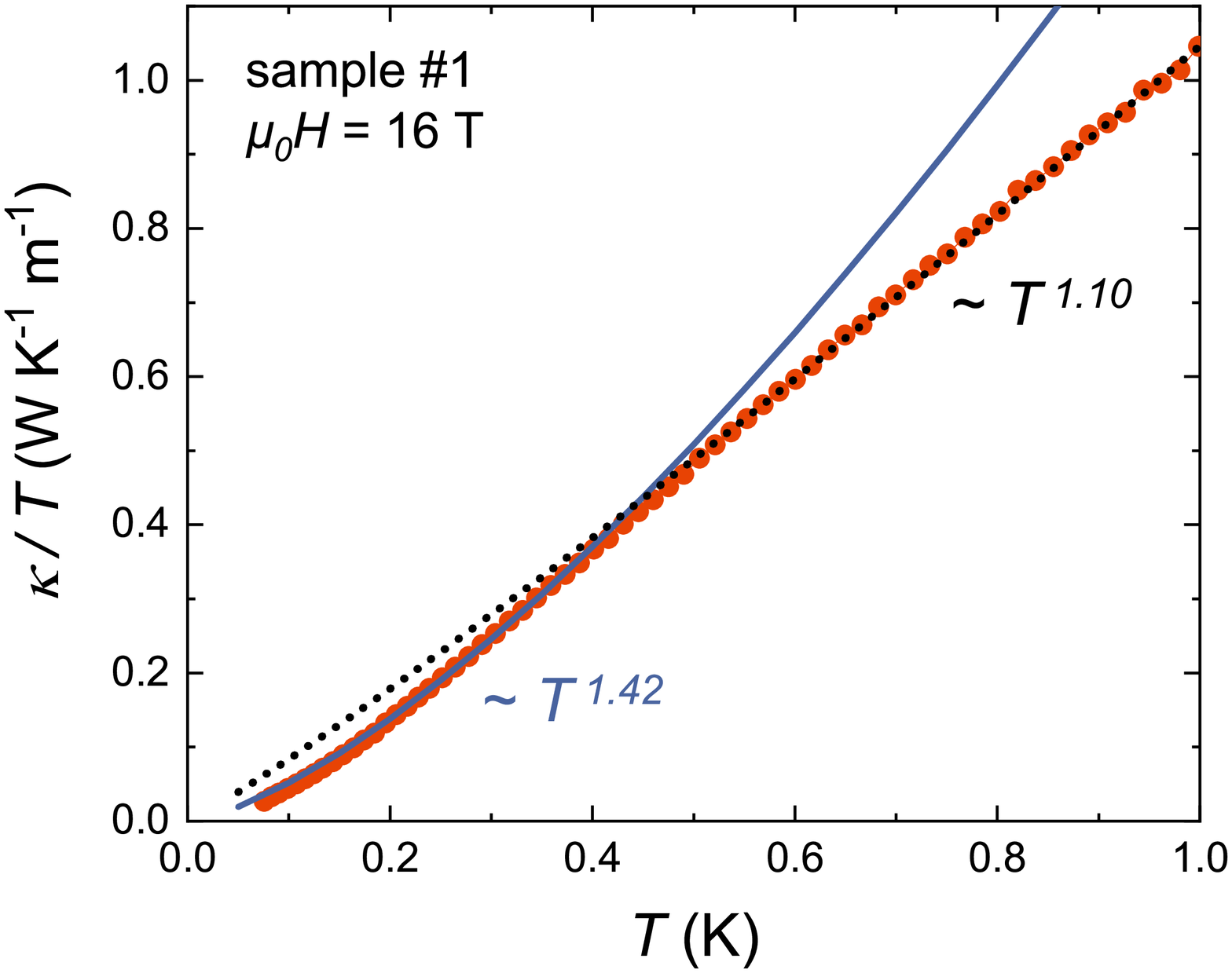}
		\caption{Thermal conductivity at \SI{16}{\tesla} of sample 1 fitted with a power law ($\kappa = C_0 \cdot T^{\alpha}$) where $\alpha = 2.42$ for $T< \SI{0,4}{\kelvin}$ (solid blue line) and $\alpha = 2.10$ for $\SI{0,4}{\kelvin} < T < \SI{1}{\kelvin}$ (dotted black line) \label{fig:kappa_fit_16T}}
	\end{figure}

\section{Exponents of the power law behavior of $l_{\text{spin-phonon}}$}

\begin{table}
	\begin{tabular}{|c|c||c|c|}
		\hline
		$\mu_0 H$ (T) & $p$ & $\mu_0 H$ (T)  & $p$ \\
		\hline \hline
		3	& -3.8	& 8.5	& -2.5  \\
		\hline
		4	& -4	& 9		&  -3.2\\
		\hline
		6	& -1.5	& 9.5	&  0.1\\
		\hline
		7	&  -4	& 10	&  -1\\
		\hline
		8.1	&  	-2.5	&  10.5	& -2.5 \\
		\hline
	\end{tabular}
	\caption{Exponents of the power law behavior of $l_{\text{spin-phonon}} = a_0 T^p$ as indicated by the dashed and solid black lines in figure \ref{fig:PhononMFP} (b).\label{tab:powerlaw}}
\end{table}

\newpage
\bibliography{LinaritePaper}
\end{document}